\documentclass[fleqn,12pt,twoside]{article}
\usepackage{espcrc1}

\def\bq{\begin{eqnarray}}
\def\eq{\end{eqnarray}}
\def\be{\begin{eqnarray}}
\def\ee{\end{eqnarray}}
\def\bra{\langle }
\def\ket{\rangle }

\usepackage{epsfig}

\newcommand{\AmS}{{\protect\the\textfont2
  A\kern-.1667em\lower.5ex\hbox{M}\kern-.125emS}}

\title{
Generalized parton distributions of hadrons
with composite constituents
}

\author{S. Scopetta\address[MCSD]{
Dipartimento di Fisica, Universit\`a degli Studi
di Perugia, via A. Pascoli, 06100 Perugia, Italy
and INFN, sezione di Perugia}}
       
\begin{document}

\maketitle

\begin{abstract}
A method is described for calculating the Generalized Parton Distributions
(GPDs) of spin 1/2 hadrons made of composite constituents,
in an Impulse Approximation framework. GPDs are obtained from the convolution
between the light cone non-diagonal momentum distribution of the hadron
and the GPD of the constituent. DIS structure functions and electromagnetic
form factors are consistently recovered with the proposed formalism.
Results are presented for the nucleon and for the $^3$He nucleus.
For a nucleon assumed to be made of composite constituent quarks,
the proposed scheme permits to study the so-called
Efremov-Radyushkin-Brodsky-Lepage (ERBL) region, difficult to access in
model calculations. Results are presented for both helicity-independent
and helicity-dependent GPDs. For $^3$He, the calculation
has been performed by evaluating a non-diagonal spectral function
within the AV18 interaction. It turns out that a measurement of GPDs for $^3$He
could shed new light on the short-range nuclear structure at
the quark level.
\end{abstract}

\vskip 0.5cm


Generalized Parton Distributions (GPDs) \cite{first}
represent one of the most relevant issues
in nowadays hadronic Physics 
(for recent reviews, 
see, e.g., Ref. \cite{dpr}).
GPDs
parameterize the long-distance dominated part of
exclusive lepton Deep Inelastic Scattering
(DIS) off hadrons and can be measured in
Deeply Virtual Compton Scattering (DVCS),
i.e. the process
$
e H \longrightarrow e' H' \gamma
$ when
$Q^2 \gg m_H^2$, permits to access GPDs
(here and in the following,
$Q^2$ is the momentum transfer between the leptons $e$ and $e'$,
and $\Delta^2$ the one between the hadrons $H$ and $H'$) so that
relevant experimental DVCS programs are taking place.
The issue of measuring GPDs for nuclei
is also being addressed \cite{bar}, following
an observation firstly discussed in \cite{cano1}. 
As a mater of facts,
the knowledge of GPDs would permit the study
of the short light-like distance structure of nuclei, 
and thus the interplay
of nucleon and parton degrees of freedom in the  
nuclear wave function.
In DIS off a nucleus
with four-momentum $P_A$ and $A$ nucleons of mass $M$,
this information can be accessed in the 
region where
$A x_{Bj} \simeq { Q^2 / ( 2 M \nu) }>1$,
being $x_{Bj}= Q^2 / ( 2 P_A \cdot q )$ and $\nu$
the energy transfer in the laboratory system.
In this kinematical region measurements are difficult, because of 
the vanishing of the cross-sections.
As explained in Ref. \cite{cano1}, 
the same physics can be accessed
in DVCS at lower values of $x_{Bj}$, 
as it will be clear later
\cite{cano2}. 
\begin{figure}[ht]
\centerline{\epsfxsize=3.0in\epsfbox{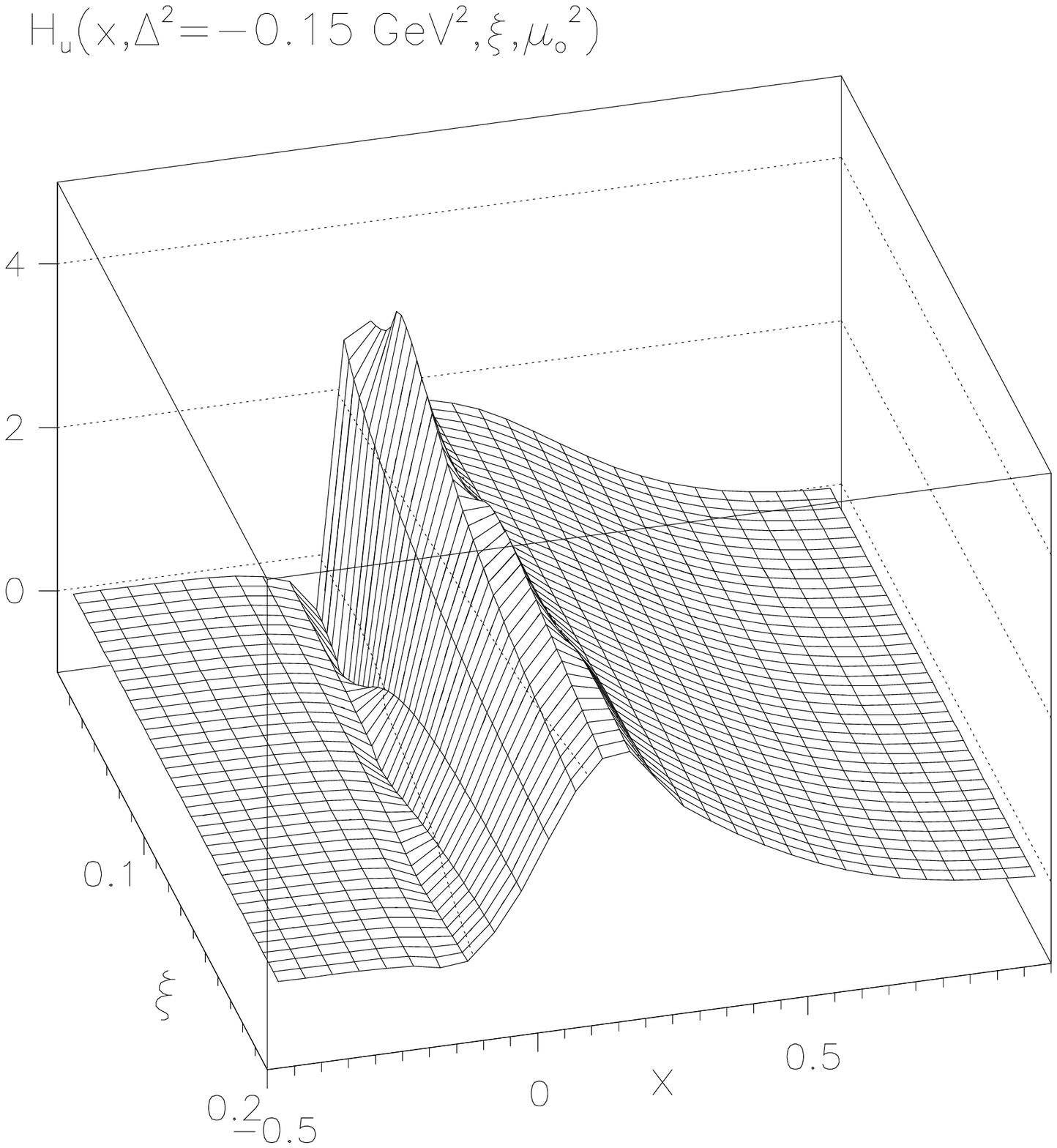}
\epsfxsize=3.0in\epsfbox{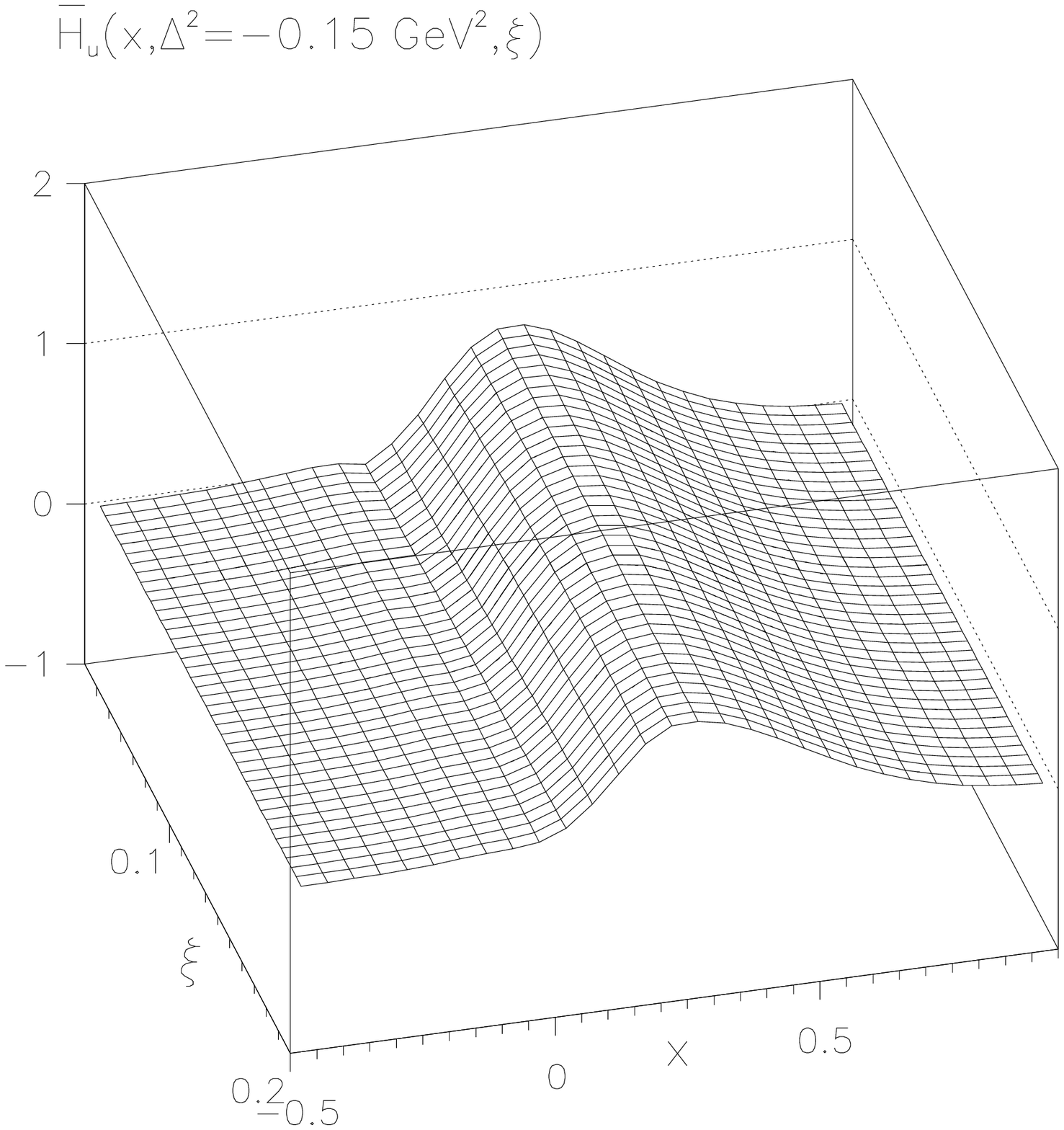}}   
\caption{
Left panel: The helicity-independent 
Non-Singlet GPD $H(x,\xi,\Delta^2)$ for the flavor $u$,
at the hadronic scale $\mu_0^2=0.34$ GeV$^2$, for $\Delta^2=-0.15$ GeV$^2$
and the allowed values of $\xi$.
Right panel: the same is shown for
the helicity-dependent GPD $\tilde{H}_u(x,\xi,\Delta^2)$.}
\end{figure}

In this talk,
a method is reviewed for calculating the 
GPDs of spin 1/2 hadrons made of composite constituents,
in an Impulse Approximation (IA) framework. 
In this scheme, GPDs are given by the convolution
between the light cone non-diagonal momentum distribution of the hadron
and the GPD of the constituent. 
Results are presented for the nucleon and for the $^3$He nucleus.
Details for the description of the nucleon target have
to be found in Refs. \cite{io3,io4}. Results are shown
here in a new kinematical region. For the $^3$He nucleus,
details can be found in Ref. \cite{prc}. Part 
of the discussion of the $^3$He case
has been presented also in \cite{gron} although the
kinematical region presented in that paper is extended
here to higher values of the momentum transfer.

If one thinks to a spin $1/2$ hadron target, with initial (final)
momentum and helicity $P(P')$ and $s(s')$, 
respectively, two 
GPDs $H_q(x,\xi,\Delta^2)$ and
$E_q(x,\xi,\Delta^2)$, occur.
If one works in a system of coordinates where
the photon 4-momentum, $q^\mu=(q_0,\vec q)$, and $\bar P=(P+P')/2$ 
are collinear along $z$,
$\xi$ is the so called ``skewedness'', defined
by the relation 
$
\xi = - {n \cdot \Delta / 2} = - {\Delta^+ / 2 \bar P^+}
= { x_{Bj} /( 2 - x_{Bj}) } + {{O}} \left ( {\Delta^2 / Q^2}
\right ) ~,
$
where $n$
is a light-like 4-vector
satisfying the condition $n \cdot \bar P = 1$.
One should notice that the variable $\xi$
can be completely fixed experimentally.
The well known constraints of $H_q(x,\xi,\Delta^2)$ are: 

i) the so called
``forward'' limit, 
$P^\prime=P$, i.e., $\Delta^2=\xi=0$, where one 
recovers the usual parton distribution
$
H_q(x,0,0)=q(x)~;
\label{i)}
$

ii)
the integration over $x$, giving the contribution
of the quark of flavor $q$ to the Dirac 
form factor (f.f.) of the target:
$
\int dx H_q(x,\xi,\Delta^2) = F_1^q(\Delta^2)~;
\label{ii)}
$

iii) the polynomiality property,
involving higher moments of GPDs, according to which
the $x$-integrals of $x^nH^q$ and of $x^nE^q$
are polynomials in $\xi$ of order $n+1$.

In Ref. \cite{io3},
an IA expression
for $H_q(x,\xi,\Delta^2)$ of a given hadron target $A$
has been obtained.
Assuming that the interacting parton belongs
to a bound constituent $N$ of the target 
with momentum $p$ and removal energy $E$,
for small values of $\xi^2$ and
$\Delta^2 \ll Q^2,M^2$, it reads:

\begin{eqnarray}
\label{flux}
H_q^A(x,\xi,\Delta^2) 
& =  & 
\sum_N \int dE \int d \vec p
\, 
P_{N}^A(\vec p, \vec p + \vec \Delta, E )
{\xi' \over \xi}
H_{q}^N(x',\xi',\Delta^2)~.
\label{spec}
\end{eqnarray}
In the above equation, the kinetic energies of the recoiling
residual system and of the recoiling target have been neglected, 
$P_{N}^A (\vec p, \vec p + \vec \Delta, E )$ is
the one-body off-diagonal spectral function
for the constituent $N$ in the target $A$,
the quantity
$
H_q^N(x',\xi',\Delta^2)
$
is the GPD of the bound constituent $N$
up to terms of order $O(\xi^2)$, and in the equation (\ref{spec})
use has been made of
the relations
$
\xi'  =  - \Delta^+ / 2 \bar p^+~,
$
and $ x' = (\xi' / \xi) x$~.
Eq. (\ref{spec}) can be written in the form
\begin{eqnarray}
H_{q}^A(x,\xi,\Delta^2) =  
\sum_N \int_x^1 { dz \over z}
h_N^A(z, \xi ,\Delta^2 ) 
H_q^N \left( {x \over z},
{\xi \over z},\Delta^2 \right)~,
\label{main}
\end{eqnarray}
where 
$ h_N^A(z, \xi ,\Delta^2 ) =  
\int d E
\int d \vec p
\, P_N^A(\vec p, \vec p + \vec \Delta) 
\delta \left( z + \xi  - { p^+ / \bar P^+ } \right)~.
$

In Ref. \cite{io3,io4}, it is discussed that
Eq. (\ref{spec}) fulfills the constraints $i)-iii)$ listed above.

The above formalism has been applied in Ref \cite{io3,io4} to
the nucleon target. The spectral function of the composite constituent
quarks has been approximated by a momentum distribution, calculated
within the Isgur and Karl model \cite{ik}, as shown in Ref. \cite{io5},
convoluted with the GPDS of the constituent quarks themselves.
The latter are modeled by using the structure functions of the constituent
quark, obtained generalizing to the GPDs case the approach of \cite{io6}
which is, in turn, built following the idea of Ref \cite{acmp},
the double distribution representation on GPDs (see, i.e., \cite{dpr,rad1}),
and a recently proposed phenomenological constituent quark 
form factor \cite{sim}.
Results have been discussed in Ref. \cite{io3} for the 
helicity-independent GPD $H(x,\xi,\Delta^2$. 
The model has been built to be valid
at the so-called hadronic scale, $\mu_0^2=0.34$ GeV$^2$,
and in Ref. \cite{io3} also
the NLO QCD evolution of the results up to typical experimental values
has been discussed and shown.
In Ref. \cite{io4} everything
has been extended to study 
the helicity-dependent GPD $\tilde H(x,\xi,\Delta^2)$,
getting a convolution involving helicity-dependent momentum
distributions. Typical results
are shown in Fig.1 for the two cases under investigation, in
a kinematical scenario which extends the one discussed in Refs.
\cite{io3,io4}.

This phenomenological approach permits to access, in a simple and physical
way, also the so-called ERBL region,
difficult to study within constituent quark model calculations.
Recently, another model approach has been proposed,
adding a meson cloud to a light-front quark model scenario 
introduced in a series of previous papers, starting with Ref.
\cite{bpt}. Such an approach leads to convolution formulas for the GPDs
and the ERBL region is also accessed through the meson cloud. 
In the latter framework
the helicity-independent GPDs have been calculated \cite{pb}.

The study of GPDs for $^3$He is interesting
for many aspects. First of all,
$^3$He is a well known nucleus, for which realistic studies 
are possible, so that conventional effects
can be calculated and the exotic ones
can be distinguished.
Besides, $^3$He is widely used as an effective 
polarized free neutron target \cite{friar} and
it will be the first candidate
for experiments aimed at the study of
GPDs of the free neutron, to unveil its angular momentum
content. 
\begin{figure}[ht]
\centerline{\epsfxsize=3.0in\epsfbox{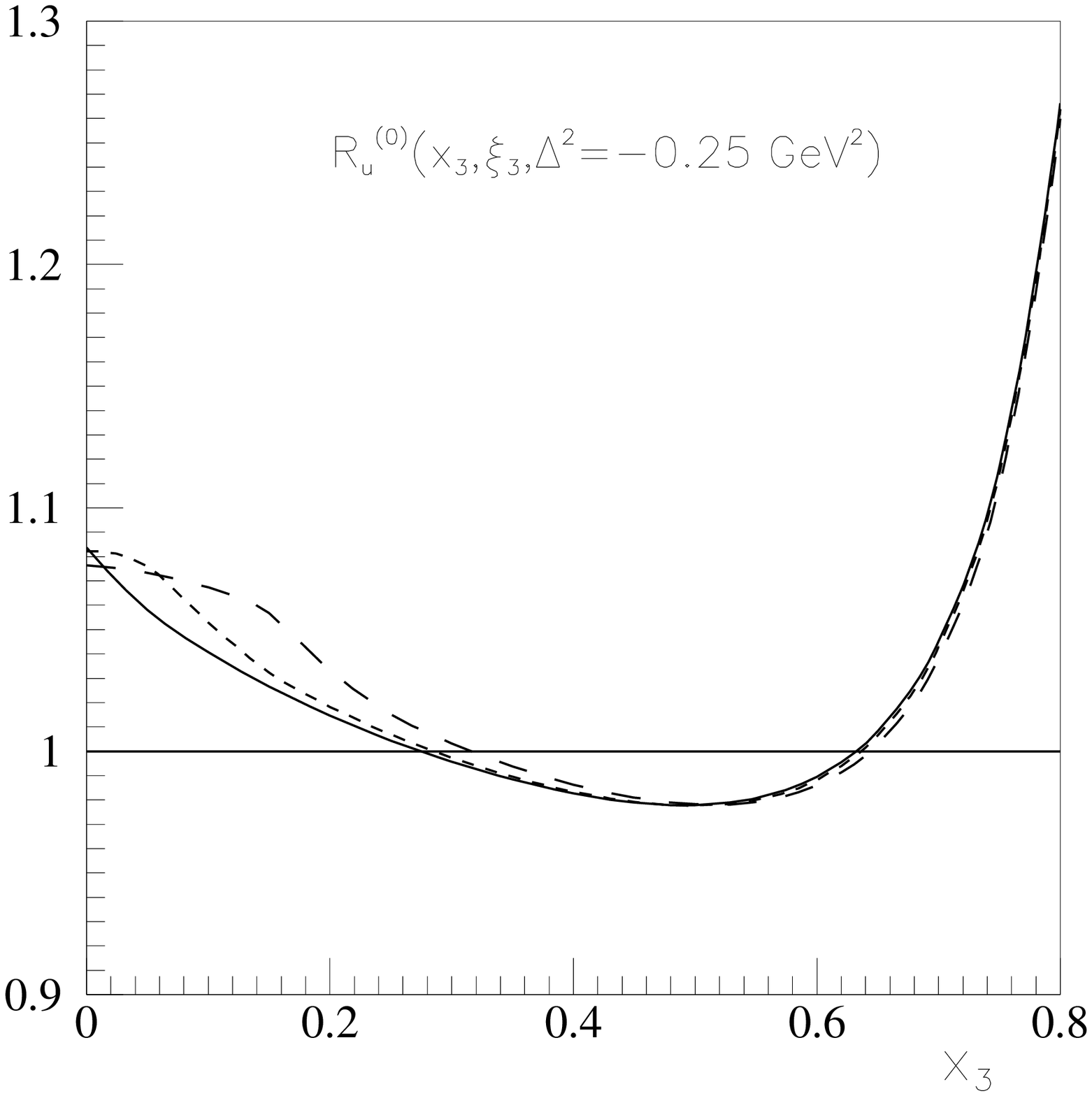}
\epsfxsize=3.0in\epsfbox{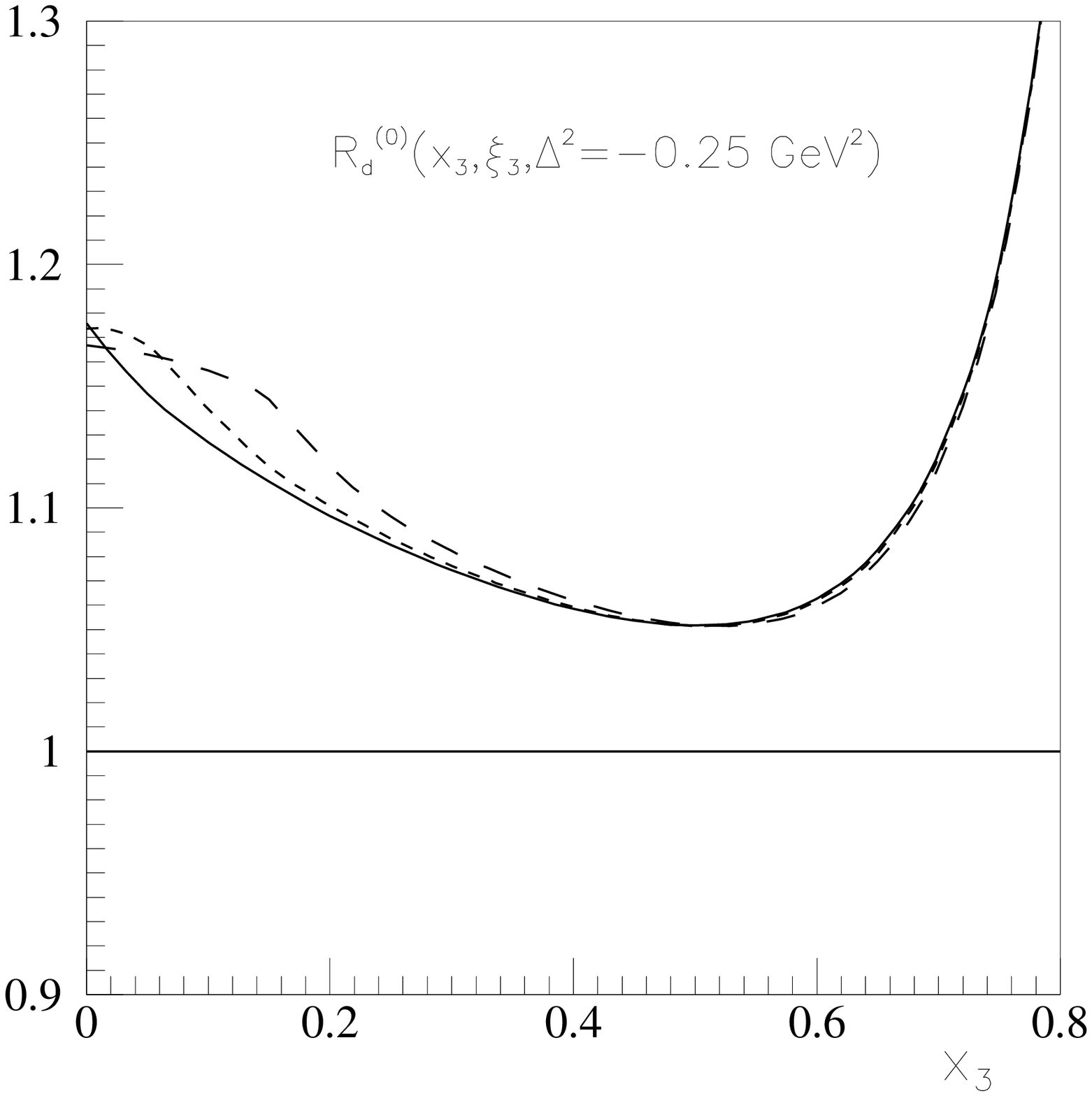}}   
\caption{
The ratio Eq. (\ref{rnew}) is shown, for
$\Delta^2 = -0.25$ GeV$^2$, 
as a function of $x_3$.
The full line has been calculated for $\xi_3=0$,
the dashed line for $\xi_3=0.1$ and the long-dashed
one for $\xi_3=0.2$. 
The symmetric part at $ x_3 \leq 0$ is not presented.
Results for the flavors $u$ and $d$ are shown in 
the left and right panels, respectively.}
\end{figure}

In what follows,
the results of an impulse approximation (IA)
calculation \cite{prc} of the unpolarized GPD $H_q^3$ 
for the quark of flavor $q$ of
$^3$He will be reviewed. A convolution formula
is discussed and evaluated
using a realistic non-diagonal spectral function,
so that Fermi motion and binding effects are rigorously estimated.
The proposed scheme is valid for 
$\Delta^2 \ll Q^2,M^2$
and despite of this it permits to
calculate GPDs in the kinematical range relevant to
the coherent channel of DVCS off $^3$He.
In fact, the latter channel is the most interesting one for its 
theoretical implications, but it can be hardly seen at
large $\Delta^2$, due to the vanishing cross section.
The 
main result of 
this investigation
is not the size and shape of the obtained $H_q^3$ for $^3$He,
but the size and nature of nuclear effects on it.
This 
permits to test, for the $^3$He  target, 
the accuracy of prescriptions
which have been proposed to estimate nuclear GPDs \cite{cano2}.




$H_q^A(x,\xi,\Delta^2)$ for $A=^3$He, Eq. (\ref{spec}), 
has been evaluated in the nuclear Breit Frame.
The non-diagonal spectral function appearing in
Eq. (\ref{spec}) has been calculated 
along the lines of Ref. \cite{gema},
by means of 
realistic wave functions 
evaluated
using the 
AV18 interaction and
taking into account
the Coulomb repulsion between the protons.
The one-body off-diagonal spectral function
for the nucleon $N$ in $^3He$ reads
\begin{eqnarray}
P_N^3(\vec p, \vec p + \vec \Delta, E)  & = & 
{1 \over (2 \pi)^3} {1 \over 2} \sum_M 
\sum_{R,s}
\bra \vec P'M | (\vec P - \vec p) S_R, (\vec p + \vec \Delta) s\ket 
\bra (\vec P - \vec p) S_R,  \vec p s| \vec P M \ket
\times
\nonumber
\\
& \times &  
\, \delta(E - E_{min} - E^*_R)~.
\label{spectral}
\end{eqnarray}
The delta function in Eq (\ref{spectral})
defines $E$, the so called removal energy, in terms of
$E_{min}=| E_{^3He}| - | E_{^2H}| = 5.5$ MeV and
$E^*_R$, the excitation energy 
of the two-body recoiling system.
The main quantity appearing in the definition
Eq. (\ref{spectral}) is
the overlap integral
\bq
\bra \vec P M | \vec P_R S_R, \vec p s \ket=
\int d \vec y \, e^{i \vec p \cdot \vec y}
\bra \chi^{s},
\Psi_R^{S_R}(\vec x) | \Psi_3^M(\vec x, \vec y) \ket~,
\label{trueover}
\eq 
between the eigenfunction 
$\Psi_3^M$ 
of the ground state
of $^3He$, with eigenvalue $E_{^3He}$ and third component of
the total angular momentum $M$, and the
eigenfunction $\Psi_R^{S_R}$, with eigenvalue
$E_R = E_2+E_R^*$ of the state $R$ of the intrinsic
Hamiltonian pertaining to the system of two interacting
nucleons.
Since the set of the states $R$ also includes
continuum states of the recoiling system, the summation
over $R$ involves the deuteron channel and the integral
over the continuum states.

The other ingredient in Eq. (\ref{spec}), i.e.
the nucleon GPD $H_q^N$, has been modelled in agreement with
the Double Distribution representation \cite{rad1}.
In this model, whose details are summarized in Ref. \cite{prc},
the $\Delta^2$-dependence of $H_q^N$ is given by
$F_q(\Delta^2)$, i.e. the contribution
of the quark of flavor $q$
to the nucleon form factor.
Now the numerical results
will be discussed.
If one considers
the forward limit of the ratio
\bq
R_q (x,\xi,\Delta^2) = 
{ H_q^3(x,\xi,\Delta^2) 
/ ( 2 H_q^p(x,\xi,\Delta^2) + H_q^n(x,\xi,\Delta^2) )}~,
\label{rat}
\eq
where the denominator clearly represents
the distribution of the
quarks of flavor $q$ 
in $^3$He if nuclear effects are completely
disregarded, 
the behavior which is found, shown in Ref. \cite{prc},
is typically $EMC-$like,
so that, 
in the forward limit, well-known results are recovered.
In Ref. \cite{prc} it is also shown that
the $x$ integral of the nuclear GPD gives a good 
description of ff data of $^3$He, in the relevant kinematical region,
$-\Delta^2 \leq 0.25$ GeV$^2$.
Let us now discuss the nuclear effects.
The full result for the GPD $H_q^3$, Eq. (\ref{spec}),
will be compared with a prescription
based on the assumptions
that nuclear effects are neglected
and the $\Delta^2$ dependence can be
described 
by
the f.f. of $^3$He:
\bq
H_q^{3,(0)}(x,\xi,\Delta^2) 
= 2 H_q^{3,p}(x,\xi,\Delta^2) + H_q^{3,n}(x,\xi,\Delta^2)~,
\label{app0}
\eq
where 
$
H_q^{3,N}(x,\xi,\Delta^2)=  
\tilde H_q^N(x,\xi)
F_q^3 (\Delta^2)
$
represents the effective GPD corresponding to the flavour $q$
of the bound nucleon 
$N=n,p$ in $^3$He. Its $x$ and $\xi$ dependences, given by the function
$\tilde H_q^N(x,\xi)$, 
is the same of the GPD of the free nucleon $N$,
while its $\Delta^2$ dependence is governed by the
contribution of the quark of flavor $q$ to the
$^3$He f.f., $F_q^3(\Delta^2)$.

The effect of Fermi motion
and binding can be emphasized showing 
the ratio
\be
R_q^{(0)}(x,\xi,\Delta^2) = { H_q^3(x,\xi,\Delta^2) / H_q^{3,(0)}
(x,\xi,\Delta^2)} 
\label{rnew}
\eq
i.e. the ratio
of the full result, Eq. (\ref{spec}),
to the approximation Eq. (\ref{app0}).
The latter is evaluated by means of the model nucleon GPDs used
as input in the calculation, and taking
$
F_u^3(\Delta^2) = {10 \over 3} F_{ch}^{3}(\Delta^2)~,
$
$
F_d^3(\Delta^2) = -{4 \over 3} F_{ch}^{3}(\Delta^2)~.
$
The choice of calculating the ratio Eq. (\ref{rnew})
to show nuclear effects is a very natural one.
As a matter of facts, the forward limit of the ratio Eq. (\ref{rnew})
is the same of the ratio Eq. (\ref{rat}), yielding the
EMC-like ratio for the parton distribution $q$ and,
if $^3$He were made of free nucleons at rest,
the ratio Eq. (\ref{rnew}) would be one.
This latter fact can be immediately realized by
observing that the prescription Eq. (\ref{app0})
is exactly obtained by
placing $z=1$, i.e. imposing no Fermi motion effects
and no convolution, into Eq. (\ref{main}). 
Typical results are shown in Fig. 2, where
the ratio Eq. (\ref{rnew}) is shown
for $\Delta^2 = -0.25$ GeV$^2$ 
as a function of $x_3=3 x$,  
for three different values
of $\xi_3=3 \xi$, for the flavors $u$ and $d$.
Some general trends of the results are apparent:
i) nuclear effects, for $x_3 \leq 0.7$, are as large as 15 \% at most;
ii) Fermi motion and binding have their main effect
for $x_3 \leq 0.3$, at variance with what happens
in the forward limit;
iii) nuclear effects increase with
increasing $\xi$ and
$\Delta^2$, for $x_3 \leq 0.3$;
iv) nuclear effects for the $d$ flavor are larger than
for the $u$ flavor.
The behaviour described above is discussed and explained
in Ref. \cite{prc}.
In general, it is found that the realistic calculation
yields a rather different result with respect to
a simple parameterizations
of nuclear GPDs, as some of the ones proposed
in Ref. \cite{cano2}.
In Ref. \cite{gron}, where a part of the
material discussed here has been presented
for the $^3$He target, it is shown that nuclear effects
are found to 
depend also on the choice of the NN potential, 
at variance with what happens in the forward case.
The study of nuclear GPDs turns out therefore to be very 
fruitful, being able to detect
relevant details of nuclear structure at short light-cone distances.
The obtained $^3$He GPDs are being used to 
estimate cross-sections in order to establish
the feasibility of experiments. A natural extension of
the proposed formalism and analysis
is the investigation of hadron helicity-flip GPDs,
which allows to study the possibility
of unveiling the quark orbital angular momentum 
contribution to the free neutron spin.

\end{document}